# Attractor of Smale-Williams type in autonomous distributed system

**V.P. Kruglov**[1,3], **S. P. Kuznetsov**[1,2] and **A. Pikovsky**[3]

[1] Saratov State University, Astrakhanskaya str., 83, Saratov, 410012, Russian Federation

[2] Kotel'nikov's Institute of Radio-Engineering and Electronics of RAS, Saratov Branch, Zelenaya str., 38, Saratov, 410019, Russian Federation

[3] Department of Physics, Potsdam University, 14469 Potsdam, Germany

We consider an autonomous system of partial differential equations for one-dimensional distributed medium with periodic boundary conditions. Dynamics in time consists of alternating birth and death of patterns with spatial phases transformed from one stage of activity to another by the doubly expanding circle map. So, attractor in the Poincaré section is uniformly hyperbolic, a kind of Smale-Williams solenoid. Finite-dimensional models are derived as ordinary differential equations for amplitudes of spatial Fourier modes (the 5D and 7D models). Correspondence of the reduced models to the original system is demonstrated numerically. Computational verification of the hyperbolicity criterion is performed for the reduced models: the distribution of angles of intersection for stable and unstable manifolds on the attractor is separated from zero, i.e. the touches are excluded. The considered example gives a partial justification to the old hopes that chaotic behavior of autonomous distributed systems may be associated with uniformly hyperbolic attractors.

**Key words:** Smale – Williams solenoid, hyperbolic attractor, chaos, Swift-Hohenberg equation, Lyapunov exponent

## *1. Introduction*

Uniformly hyperbolic chaotic attractors such as Smale-Williams solenoid or Plykin attractor were introduced in mathematical theory of dynamical systems several decades ago [1-5]. Figure 1 shows a construction of the Smale-Williams attractor: at one step a toroidal domain transforms into a narrow tube in a form of double loop embedded inside the original domain. After many steps, asymptotically, the solenoid arises having an infinite number of turns and Cantor-like transversal structure. Once it was believed that such attractors may describe chaos and turbulence in many cases, but later it turned out that chaotic dynamics commonly occurring in applications do not fit the class of uniformly hyperbolic attractors.

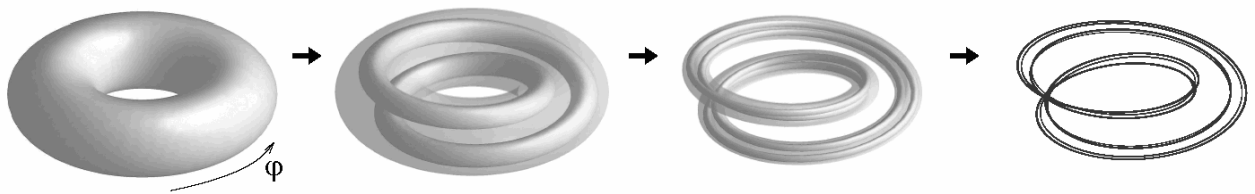

**Figure 1:** An original toroidal domain in the phase space, and results of its transformation at successive steps of discrete-time evolution with formation of the Smale-Williams solenoid after a large number of repetitive of the mapping. The angular variable φ undergoes a double expansion on each next step.

Physically implementable systems with hyperbolic chaos were discovered (or, rather constructed) only very recently [6-10]. Their principle of operation was based on chaotic nature of maps for the angular variables characterizing phases of oscillations in time, at successive stages of activity of oscillatory elements constituting the system.

An alternative general approach to elaboration of systems with hyperbolic chaos, appropriate for spatially extended systems, was advanced in Ref. [11]: instead of phases of oscillations in time, it was suggested to deal with *spatial phases* of patterns generated in a distributed medium. In computations, this principle was demonstrated and illustrated with a model based on a modified Swift-Hohenberg equation. Due to externally forced periodic modulation of a parameter controlling the characteristic spatial scale, the model system generates long-wave and short-wave Turing patterns alternately, and operates in such way that the spatial phases at successive stages of the dynamical evolution are governed by an expanding circle map. One more example with analogous dynamics of the spatial phases relates to alternating parametric excitation of long-scale and short-scale standing-wave patterns due to pump modulation in a medium described by a wave equation with nonlinear dissipation [12, 13]. Such examples justify, at least particularly, the old expectations for applicability of the hyperbolic theory to chaotic dynamics in spatially extended systems. Some dissatisfaction is the fact that these examples relate to non-autonomous systems, with time-dependent coefficients in the partial differential equations. It would be interesting to discover *autonomous* extended systems manifesting the hyperbolic chaos. Elaboration of such an example is a goal of the present article; we will present an autonomous set of nonlinear partially differential equations manifesting attractor of Smale-Williams type in the Poincaré map. At the moment we do not pretend to relate this construction to a concrete physical system, but we suppose that it may be implementable, in particular, on a base of electronic circuits (say, in a kind of nonlinear transmission line).

## 2. The main model and results of computer simulation

Following [11], let us start with the one-dimensional Swift-Hohenberg equation

$$\partial_t u + (1 + \kappa^2 \partial_x^2)^2 u = \mu u - u^3. \qquad (1)$$

A trivial solution of this equation $u \equiv 0$ becomes unstable at $\mu > 0$. In linear approximation substituting $u \sim \exp(\lambda t - ikx)$ we evaluate the increment of the perturbation growth as $\lambda = \mu - (1 - \kappa^2 k^2)^2$. It is maximal at $k = k_0 = 1/\kappa$ that corresponds roughly to the wave number of the Turing pattern, which grows up in the medium from arbitrarily small random initial perturbations. Due to the cubic nonlinear term in (1) the pattern, which is nearly periodic in space, saturates at some finite level of magnitude.

To modify the model, we add one more variable $v$ depending on the spatial coordinate $x$ and time $t$, and turn to the following set of equations

$$\begin{aligned}\partial_t u + (1 + \partial_x^2)^2 u &= \mu u + u^3 - \tfrac{1}{5} u v^2 + \varepsilon v \cos 3x, \\ \partial_t v &= -v + u^2 v + u^2.\end{aligned} \qquad (2)$$

For our purposes it is appropriate to postulate circular geometry of the medium:

$$u(x+L,t) = u(x,t),\ v(x+L,t) = v(x,t), \qquad (3)$$

where $L$ is the system length. The first relation in (2) is the modified Swift-Hohenberg equation (with inverse sign of the nonlinear term), and with added terms containing the new variable $v$ governed by the second equation. The evolution rule for the variable $v$ is local (no spatial derivatives are accounted). The term proportional to $\varepsilon \cos 3x$ implies periodic spatially inhomogeneous coupling between two components involved in the dynamics.

In accordance with the periodic boundary conditions, the functions $u$ and $v$ describing the pattern are spatially periodic and may be represented by Fourier series expansions.

Qualitatively, the functioning of the systems may be explained as follows. Suppose at some moment the multiplier $\mu + u^2 - \frac{1}{5}v^2$ in the first equation (2) is positive; hence, the variable $u$ grows, giving rise to a spatial pattern with appropriate wave-number about $k_0=1$. The most significant is the contribution of the first spatial harmonic component, which is characterized naturally by some phase $\varphi$: $u \approx U_1 \cos(x+\varphi)$. After a while, due to the growth of the pattern magnitude, the coefficient $(-1+u^2)$ in the second equation becomes positive (except narrow spatial neighborhoods of nodes of the pattern), and then the amplitude of the variable $v$ starts to grow. In the function $v$ the second spatial harmonic component is relevant, and at the initial stage of this instability it accepts the doubled spatial phase $2\varphi$ as it is stimulated by the quadratic nonlinear term $u^2$ in the second equation (2). So, we have $v \approx V_0 + V_2 \cos(2x+2\varphi)$. Next, as the magnitude of $v$ becomes large, the factor $\mu + u^2 - \frac{1}{5}v^2$ turns out to be negative. The magnitude of $u$ starts to decrease, and sequentially, the amplitude of $v$ decreases too. This process continues until the factor $\mu + u^2 - \frac{1}{5}v^2$ once more becomes positive, and the variable $u$ starts to grow again. The excitation of the first subsystem on this stage is stimulated by the term composed as a product of the second harmonic component of $v$ and the function $\varepsilon \cos 3x$ accounting the spatial inhomogeneity. It provides transfer of the doubled spatial phase back to the pattern of $u$ (with the opposite sign), as follows from the relation $\cos(2x+2\varphi)\cos 3x = \cos(x-2\varphi)/2 + \ldots$. Then, the process repeats itself over and over again. So, the spatial phases on successive stages of the pattern formation evolve according to the expanding circle map $\varphi_{n+1} = -2\varphi_n + \text{const}$. This is a chaotic map (the Bernoulli map) characterized by the positive Lyapunov exponent $\Lambda = \ln 2 = 0.693\ldots$.

Numerical simulation of the dynamics was performed at $\mu = 0.03$, $\varepsilon = 0.03$, $L = 2\pi$ using the explicit-implicit difference scheme on a grid with temporal and spatial steps, respectively, $\Delta t = 0.001$, $\Delta x = L/64 \approx 0.098$. One can observe relatively long-time stages of growth of the patterns alternating with their fast decay almost to zero (Fig. 2). The mean time period between the successive stages of excitation in this regime according to the computations is $\tau = 50.37$.

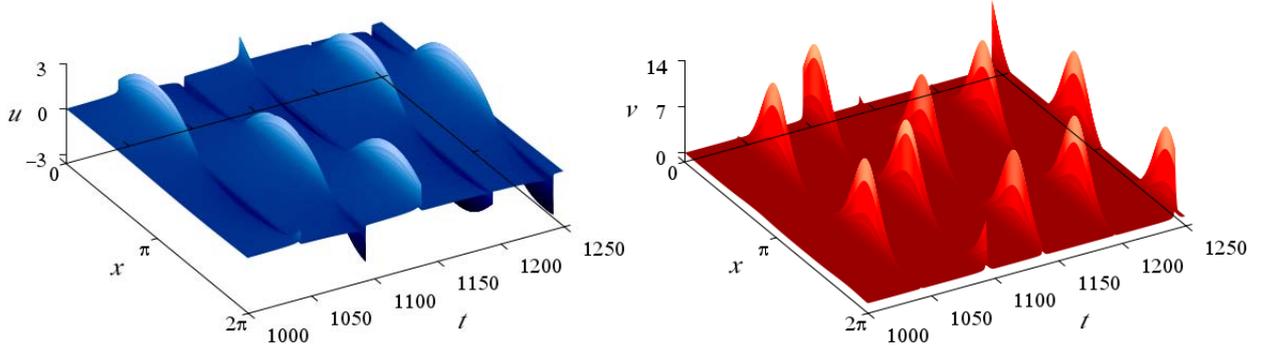

**Figure 2:** Spatio-temporal plots for the variables $u(x, t)$ (a) and $v(x, t)$ (b) obtained from numerical solution of the equations (2) with periodic boundary conditions at $\mu=0.03$, $\varepsilon=0.03$, $L=2\pi$

To analyze chaotic behavior of the patterns it is appropriate to deal with harmonic components of the spatial Fourier series expansion for $u(x, t)$ and $v(x, t)$. The complex amplitudes of the Fourier components were evaluated numerically in the course of the integration of the equations at each time step. In Fig. 3 the diagrams (a)-(f) show the amplitudes of the harmonic components, $U_1$, $U_3$ and $V_0$, $V_2$, $V_4$, $V_6$, on a time interval of ten characteristic periods of the dynamics. Note that magnitude of $U_1$ just before the drops is much larger then that of the third harmonic component $U_3$. Magnitudes of the Fourier components $V_0$, $V_2$ and $V_4$ are roughly of the same order. The even components of $u$ and odd components of $v$ are of negligible small amplitude.

An important observation is that spatial phases of the patterns on successive stages of activity evolve chaotically as seen from Figs. 2 and 3. To analyze the dynamics of the phases in the computations, we apply the Poincaré section technique. An appropriate selection for the surface of the cross-section in the state space of the system is determined by a condition that the amplitude of the first harmonic component $U_1$ passes the value 1 decreasing in time, i.e. satisfies the equality $S = |U_1| - 1 = 0$.

Figure 4 shows an iteration diagram for the spatial phases of the pattern $u(x, t)$ computed as values of argument of the complex amplitude $U_1$ at successive passages of the Poincaré section and portrait of attractor in the Poincaré section shown in two-dimensional projection onto a plane of real and imaginary parts of $U_1$. As seen from panel (a), dynamics of the spatial phase correspond to the expanding circle map (the Bernoulli map). Indeed, one complete round from 0 to $2\pi$ for the pre-image $\varphi_n$ corresponds to two rounds for the image $\varphi_{n+1}$ (in the reverse direction).

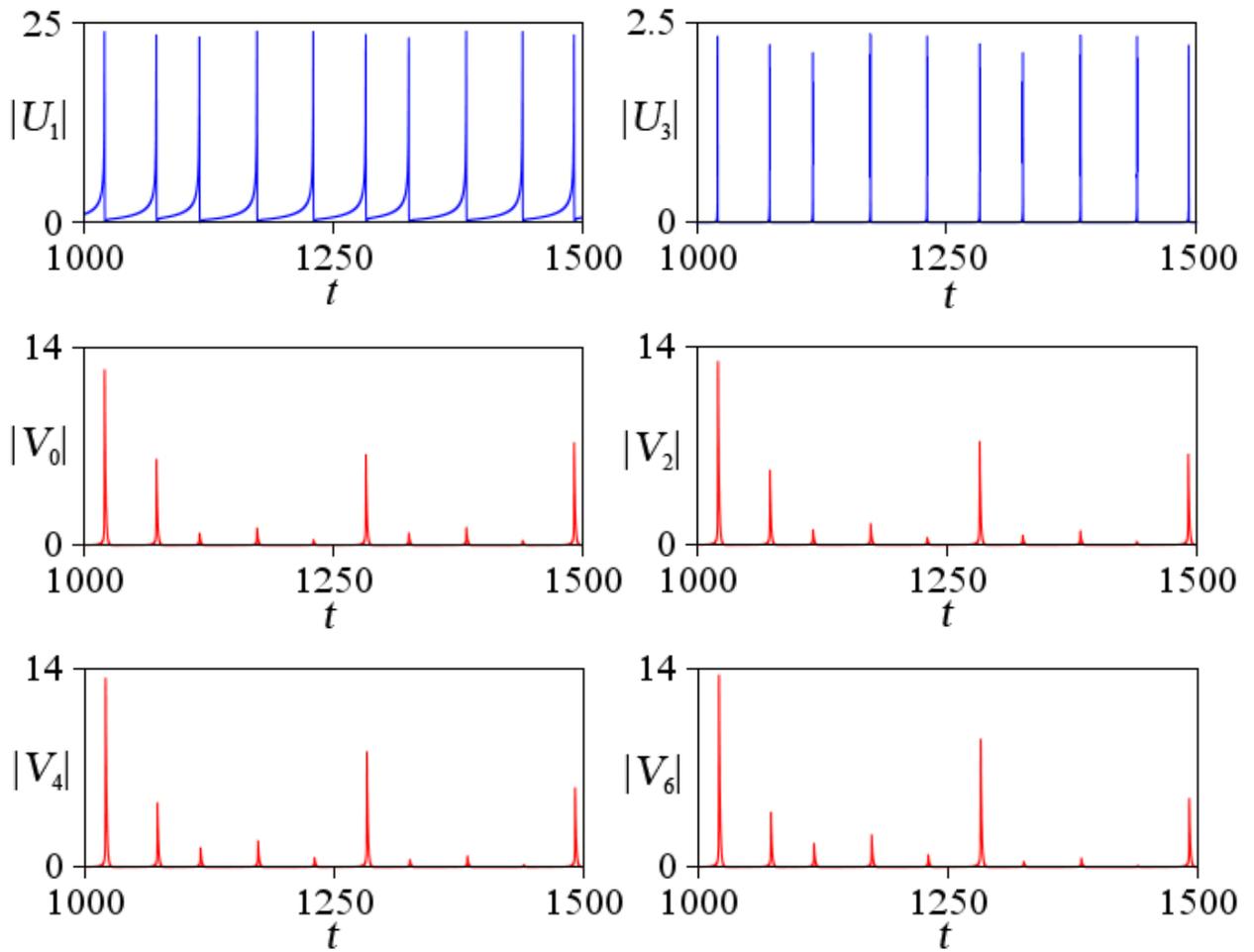

**Figure 3:** Absolute values of the amplitudes of the spatial Fourier components $U_1$, $U_3$ and $V_0$, $V_2$, $V_4$, $V_6$ for the variables $u$ and $v$ versus time as obtained from numerical solution of (2), (3) at $\mu=0.03$, $\varepsilon=0.03$, $L=2\pi$

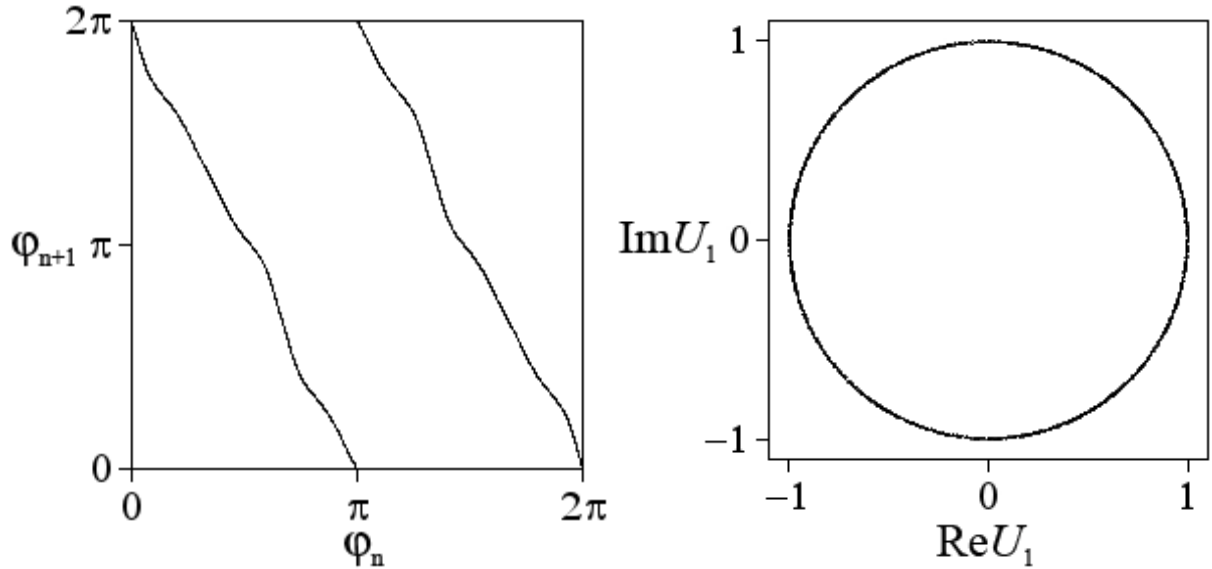

**Figure 4:** Diagram for spatial phases of the patterns for the variable $u$ evaluated as arguments of the complex amplitude of the first spatial harmonic $U_1$ at successive crossing of the Poincaré section (a) and projection of the attractor in the Poincaré section onto the plane of real and imaginary parts of $U_1$ (b) at $\mu=0.03$, $\varepsilon=0.03$, $L=2\pi$

For the sustained chaotic regime we computed Lyapunov exponents for the Poincaré map. It was done using the Benettin algorithm adapted for the distributed system [14-16, 9], with Gram-Schmidt orthogonalization of the perturbation vectors corresponding to a restricted number of modes of spatially depending variations nearby the analyzed pattern evolving in time. At $\mu = 0.03$, $\varepsilon = 0.03$, $L = 2\pi$ the first four exponents (from the infinite spectrum of those in the distributed system) are $\Lambda = \{0.665, -42.26, -44.51, -46.46, ...\}$. The largest Lyapunov exponent for the Poincaré map is remarkably close to the value ln2 corresponding to the uniformly expanding circle map. Other exponents are negative.

Summarizing, we conclude that in the state space of the system there is twofold expansion of a phase-volume element along some circular variable (that is the spatial phase of the pattern) accompanied by compression in other directions. It means that we deal with an attractor that is a kind of Smale-Williams solenoid embedded in the (infinite-dimensional) state space of the Poincaré map of our system. Estimate of the Kaplan-Yorke dimension basing on the spectrum of the Lyapunov exponents yields $D_{KY} = 1.016$. The fractional part of the dimension is rather small because of strong transversal compression of the phase volume nearby the attractor.

## *3. The 5D model*

Having in mind that the functioning of the system according to the above qualitative explanation is based on interaction of spatial Fourier harmonics, one can try to reduce the dynamics to finite-dimensional ordinary differential equations for amplitudes of the most significant modes of the components *u* and *v*.

The simplest set of shortened equations, which correctly reflects the qualitative character of the dynamics, may be derived by substitution of the following ansatz:

$$u = U_1 e^{ix} + U_1^* e^{-ix}, \quad v = w + V_2 e^{2ix} + V_2^* e^{-2ix}. \tag{4}$$

Here $U_1(t)$ is the complex amplitude of the first harmonic of the function *u*, $V_2(t)$ is the complex amplitude of the second harmonic of the function *v*, and *w*(*t*) is a real variable corresponding to the spatially-independent component of *v*. With this substitution in (2), multiplying the first equation by $e^{-ix}$, and the second by $e^{-2ix}$ or 1, after averaging over a spatial period $2\pi$ we arrive at the following set of equations

$$\begin{aligned}
\dot{U}_1 &= (\mu - \tfrac{1}{5}w^2 + 3|U_1|^2 - \tfrac{2}{5}|V_2|^2)U_1 - \tfrac{2}{5}V_2 U_1^* w + \tfrac{1}{2}\varepsilon V_2^*, \\
\dot{V}_2 &= (2|U_1|^2 - 1)V_2 + (w+1)U_1^2, \\
\dot{w} &= -w + 2\operatorname{Re}(V_2 U_1^{*2}) + 2|U_1|^2 (w+1).
\end{aligned} \tag{5}$$

Figure 5 illustrates some results of numerical solution of (5) by the Runge-Kutta fourth order method. Here the time dependences are plotted for amplitudes and arguments of the complex variables $U_1$ and $V_2$ as well as that for the real variable *w*. One can see that the plots look similar to those corresponding to the original model governed by the partial differential equations. Jumps on the plots for the phases correspond to short time intervals of transfer of the excitation between the spatial modes as explained.

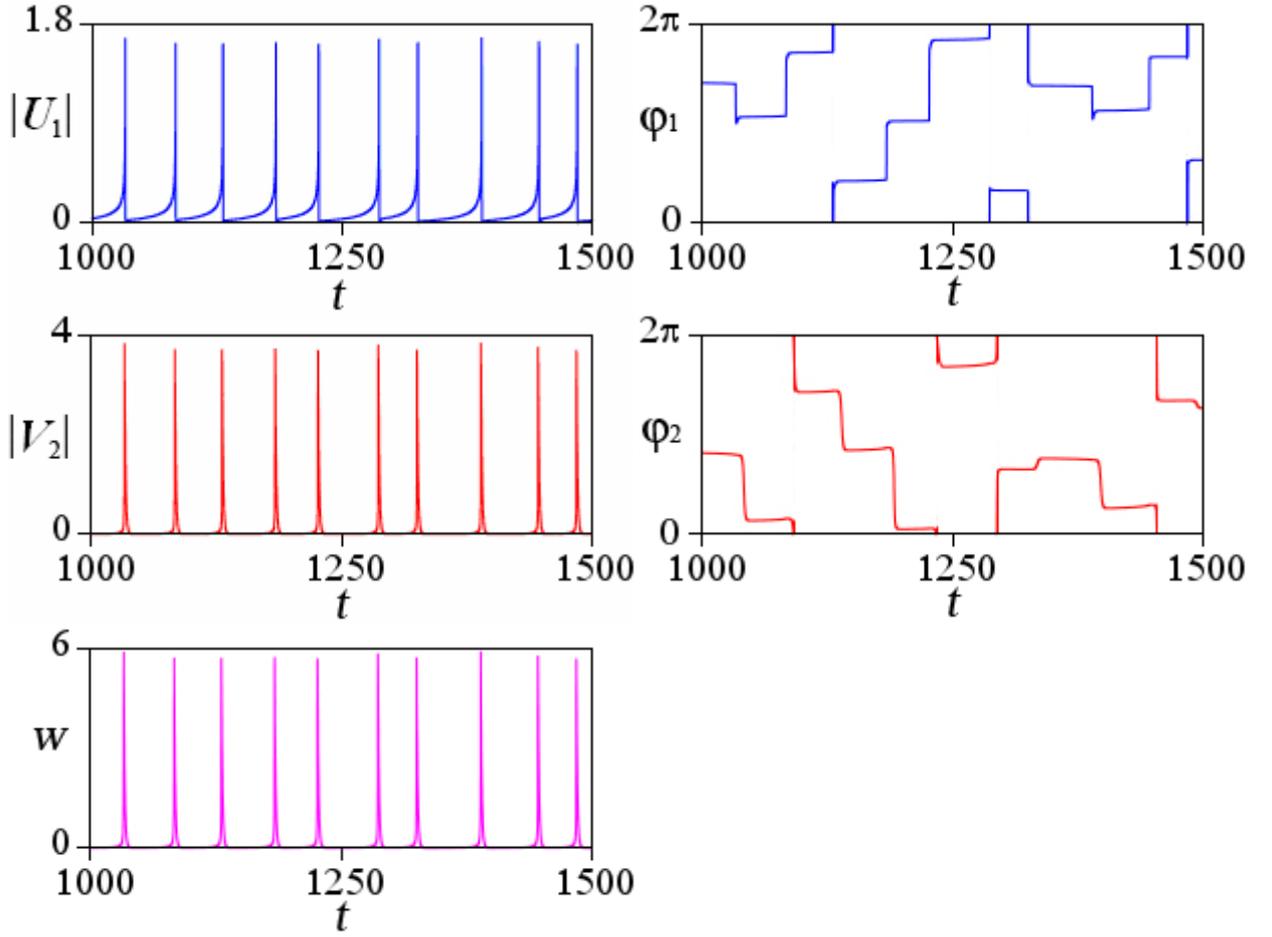

**Figure 5:** Time dependences for amplitudes and phases of the complex variables $U_1$ and $V_2$ and for the real variable $w$ obtained from numerical solution of (5) μ=0.03, ε=0.03

To construct numerically the Poincaré map we select the surface of the cross-section by the relation $S = |U_1| - 1 = 0$ and consider the variables at the instants of crossing this surface by the orbits in direction of decrease of $U_1$. For the five-dimensional autonomous set of equations (5) the Poincaré map is four-dimensional. Figure 5a shows the diagram for the phases of the variable $U_1$ at successive crossing the surface of the Poincaré section at $\mu = 0.03$, $\varepsilon = 0.03$. As seen, the dynamics of the phases in the low-dimensional model is governed by the expanding circle map for the phases, although undulating of the curve is much more pronounced than that in the original model. Figure 5b shows attractor in the Poincaré cross-section in projection onto a plane of real and imaginary parts of $U_1$. For this attractor we computed the Lyapunov exponents for the Poincaré map by the Benettin algorithm with Gram-Schmidt orthogonalization of the perturbation vectors around the reference orbit on the attractor, they are $\Lambda = \{0.65, -46.95, -49.57, -51.14\}$. The mean period of the passage between the successive crossings of the Poincaré section was found in computations to be $\tau = 52.61$. Note that the largest Lyapunov exponent is close to ln2, and others are negative. It corresponds to attractor of Smale-Williams type in the four-dimensional state space of the Poincaré map. Estimate of the fractal dimension according the Kaplan-Yorke formula yields $D_{KY} = 1.014$.

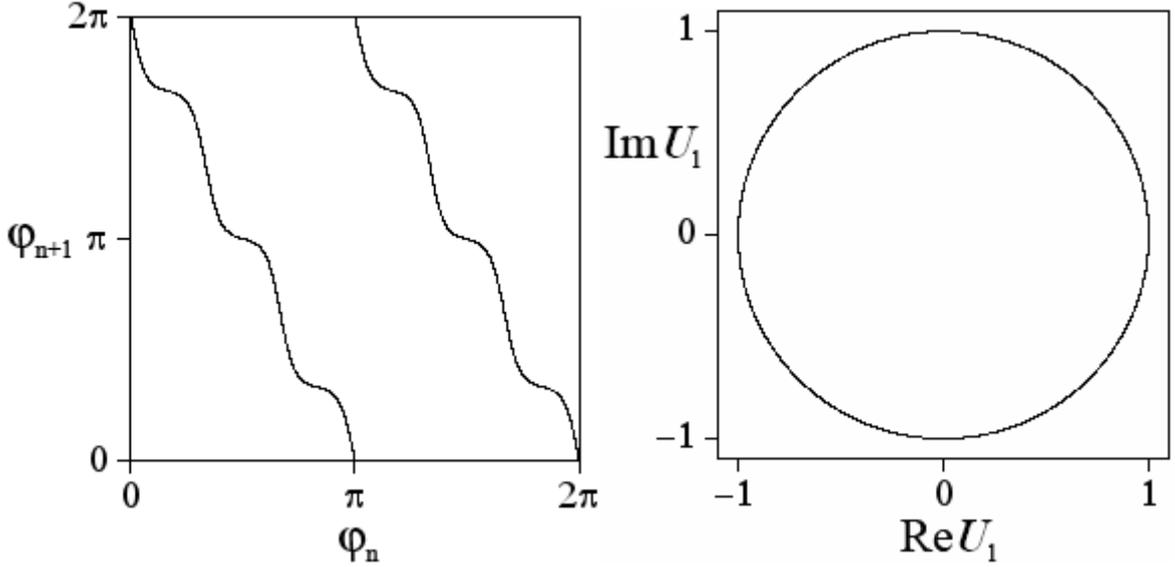

**Figure 6:** Diagram for phases of complex variable $U_1$ at successive crossing of the Poincaré section (a) and projection of the attractor in the Poincaré section onto the plane of real and imaginary parts of $U_1$ (b) for the 5D-model (5) at µ=0.03, ε=0.03

## *4. The 7D model*

One can increase the number of accounting spatial Fourier components to obtain more accurate finite-dimensional models of larger phase space dimension. The system (5) is the model with minimal number of dynamical variables to catch main features of the dynamics, but quantitatively the description becomes much better in the seven-dimensional model. Let us use instead of (4) the following substitution

$$u = U_1 e^{ix} + U_1^* e^{-ix}, \quad v = w + V_2 e^{2ix} + V_2^* e^{-2ix} + V_4 e^{4ix} + V_4^* e^{-4ix}, \quad (6)$$

where $V_4$ is the complex amplitude of the fourth spatial Fourier component of the variable $v(x, t)$. Then, from the manipulations similar to those in derivation of the model (5), we obtain the set of ordinary differential equations

$$\begin{aligned}
\dot{U}_1 &= (\mu - \tfrac{1}{5}w^2 + 3|U_1|^2 - \tfrac{2}{5}|V_2|^2 - \tfrac{2}{5}|V_4|^2)U_1 - \tfrac{2}{5}V_2 U_1^* w - \tfrac{2}{5}U_1^* V_2^* V_4 + \tfrac{1}{2}\varepsilon(V_2^* + V_4), \\
\dot{V}_2 &= (2|U_1|^2 - 1)V_2 + V_4 U_1^{*2} + (w+1)U_1^2, \\
\dot{V}_4 &= (2|U_1|^2 - 1)V_4 + U_1^2 V_2, \\
\dot{w} &= -w + 2\operatorname{Re}(V_2 U_1^{*2}) + 2|U_1|^2 (w+1).
\end{aligned} \quad (7)$$

(We do not account additional harmonics in the Fourier expansion for the function $u(x, t)$ as they appear to have small amplitudes and do not influence notably the accuracy of the description.)

Figure 7 illustrates numerical results obtained for the model (7) at $\mu = 0.03$, $\varepsilon = 0.03$; the time dependences are plotted for the amplitudes and arguments of $U_1$, $V_2$, $V_4$ and for the real variable *w*. Using the relation $S = |U_1| - 1 = 0$ and considering the variables at the instants of crossing this surface by the orbits in direction of decrease of $U_1$ we construct the Poincaré map, which is six-dimensional for this model. Figure 8a shows the diagram for the phases of $U_1$ at successive crossing the surface of the Poincaré section. It corresponds to the expanding circle map, like in the main model of Section 2 and the low-dimensional model of Section 3. Observe that the undulating in the form of the curve has decreased essentially in comparison with the five-dimensional model, and the form corresponds much better to the original distributed system (cf. Fig. 4a). Figure 8b shows attractor in the Poincaré cross-section in projection onto a plane of real and imaginary parts of the variable $U_1$.

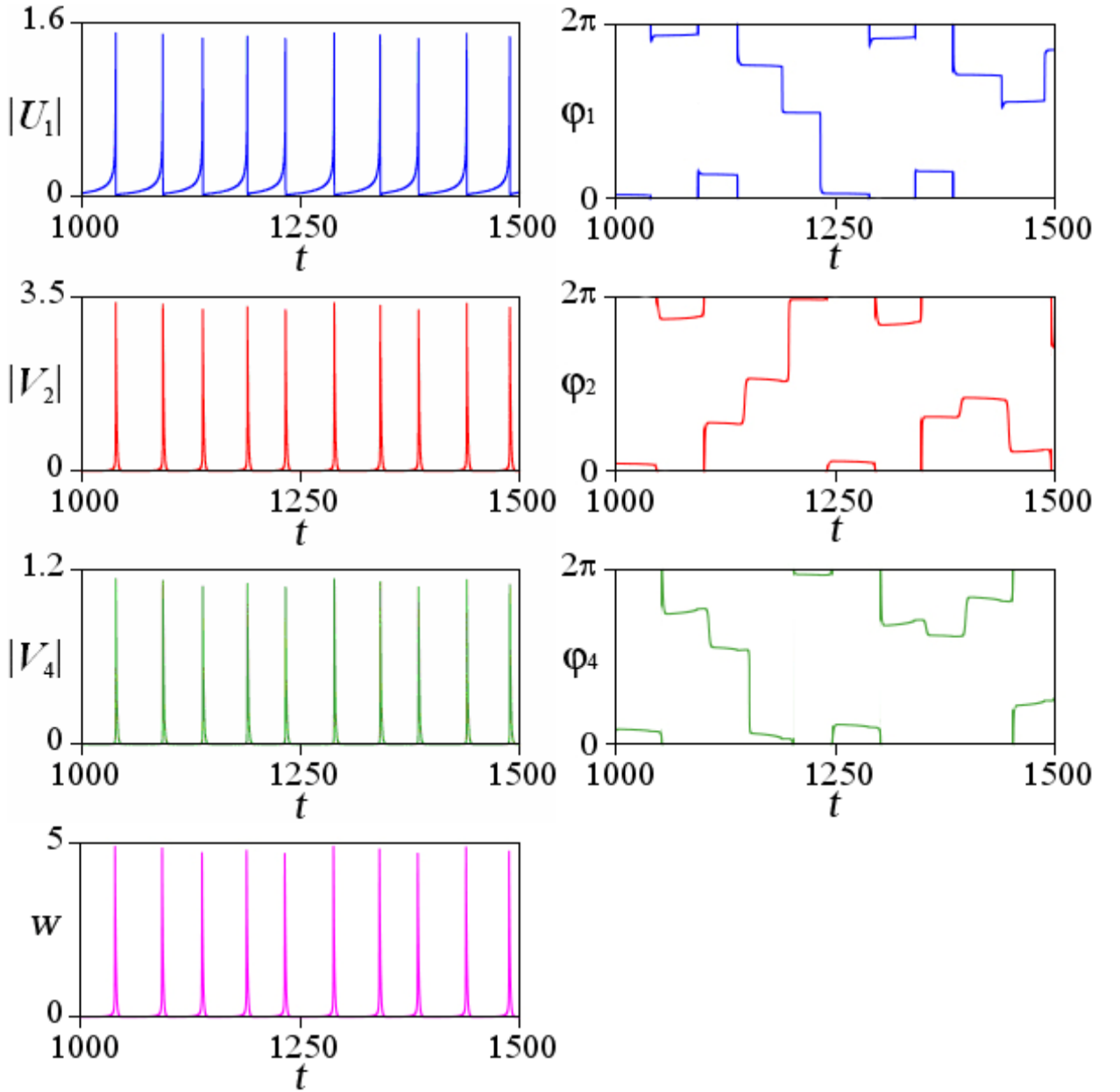

**Figure 5:** Time dependences for amplitudes and phases of the complex variables $U_1$, $V_2$, $V_4$ and for the real variable $w$ obtained from numerical solution of (7) at μ=0.03, ε=0.03

Lyapunov exponents for the Poincaré map computed by the Benettin algorithm with Gram-Schmidt orthogonalization are $\Lambda = \{0.67, -41.72, -46.34, -47.35, -47.62, -47.81\}$. Observe the better agreement with the exponents of the distributed model comparing with the five-dimensional model. The mean period of the passages between successive crossings of the Poincaré section in this regime was found to be $\tau = 49.6$. Again we see that the largest Lyapunov exponent is close to ln2, and others are negative that corresponds to attractor of Smale-Williams type embedded in the six-dimensional state space of the Poincaré map. Estimate of the fractal dimension according the Kaplan-Yorke formula yields $D_{KY} = 1.016$ that agrees with the distributed model.

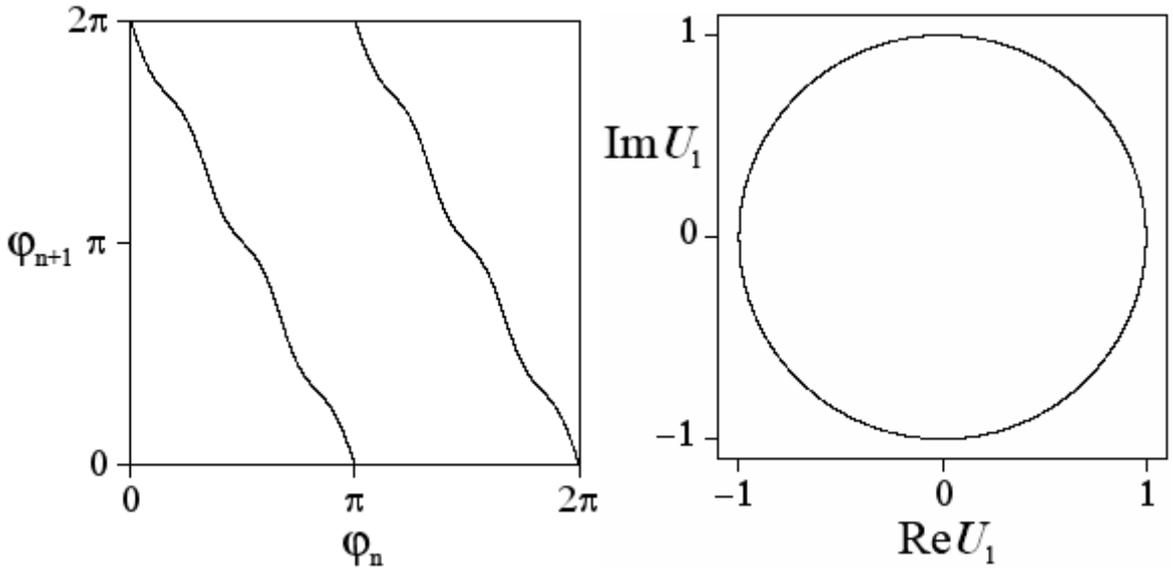

**Figure 8:** Diagram for phases of complex variable $U_1$ at successive crossing of the Poincaré section (a) and projection of the attractor in the Poincaré section onto the plane of real and imaginary parts of $U_1$ (b) for the 7D-model (7) at μ=0.03, ε=0.03

## 5. Hyperbolicity test for the finite-dimensional models

For attractors of the 5D and 7D models discussed in two previous sections, we have conducted a numerical test for hyperbolicity following the approach suggested in [17, 18] and used in application to hyperbolic attractors in [6-9].

The method is based on estimate of distribution of angles between stable and unstable manifolds on an attractor. The stable and unstable manifolds for the hyperbolic attractor can meet only with nonzero angle; the touches must be excluded. (Their presence would signalize about non-hyperbolic nature of the attractor, see examples e.g. in [9].) The procedure consists in computation of vectors of small perturbations along a representative trajectory on the attractor in forward and inverse time, with measuring angles between the forward-time vectors and the spanned subspace of vectors unstable in the backward-time at crossings of the Poincaré section by the reference trajectory. In our case the unstable manifold is one-dimensional, and the stable manifold is of dimension $N-1$, where $N$=4 or 6 is the phase space dimension for the Poincaré map. Instead of tracing all relevant vectors in the backward-time, in the computations one can deal with only one vector generated by the *conjugate* linear equations for perturbations at the reference trajectory; see details in Ref. [19], where this modification of the method was suggested. If zero values of the angle do not occur, i.e. the statistical distribution of the angles is separated from zero, one concludes that the dynamics is hyperbolic. If the statistics show non-vanishing probability for zero angles, it implies non-hyperbolic behavior because of presence of the homoclinic tangencies of the stable and unstable manifolds.

In our 5D and 7D models we generate, first, a sufficiently long representative orbit on the attractor from the numerical solution of Eqs. (5) or (7). Then, we integrate numerically the linearized variation equations forward in time to get a perturbation vector **a**($t$) normalizing it at each step of integration to exclude the divergence. This vector determines an unstable direction at each point of the orbit. Next, we solve the conjugate linearized variation equations along the same reference trajectory in backward time to get a vector **v**($t$) orthogonal to the three-dimensional stable subspace. Then, we compute an angle $\beta \in [0, \pi/2]$ between **v**($t$) and **a**($t$) from the relation $\cos\beta = |\mathbf{v}(t) \cdot \mathbf{a}(t)| / |\mathbf{v}(t)||\mathbf{a}(t)|$ and set $\alpha = \beta - \pi/2$.

Figure 9 shows histograms for distributions of the angles $\alpha$ obtained in computations for the 5D and 7D models at $\mu = 0.03$ and $\varepsilon = 0.03$. Observe clearly visible separation of the distributions from zero values of the angles. So, the test confirms hyperbolicity of the attractors.

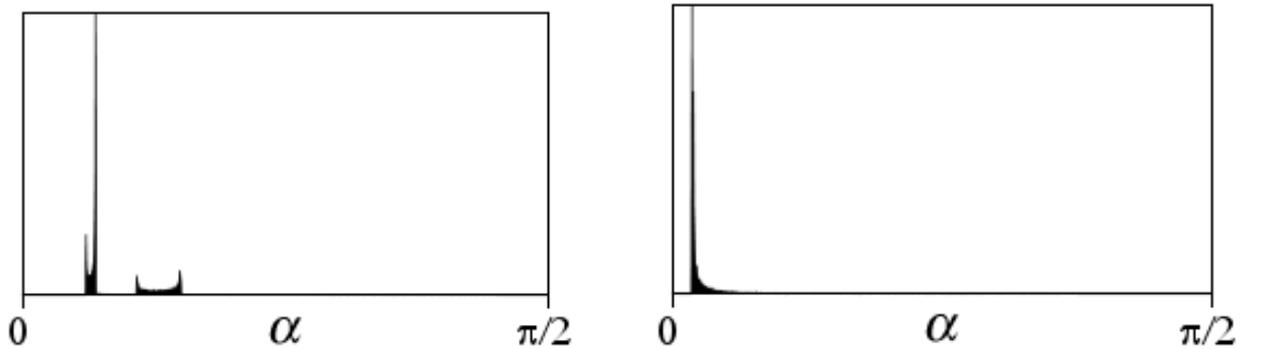

**Figure 9:** Histograms for distributions of the angles between the stable and unstable subspaces on the attractors for the models of dimension 5 (a) and 7 (b) obtained in computations at μ=0.03, ε=0.03

## *6. Conclusion*

In this article, we have discussed an example of autonomous distributed system with ring geometry (periodic boundary condition), which implements chaotic dynamics corresponding to uniformly hyperbolic attractor, a kind of Smale-Williams solenoid in the Poincaré map. The dynamics consists of sequential birth and death of the spatial patterns; the Smale-Williams attractor occurs due to the fact that the spatial phases of these patterns on each next stage of activity are transformed according to the doubly expanding circle map. Also we derived and studied numerically the truncated models represented by a five-dimensional and a seven-dimensional set of ordinary differential equations. Their dynamics are found to correspond qualitatively to the original distributed system. For these models, the hyperbolicity of the attractor is confirmed by a computer-based test, which indicates the lack of touches for stable and unstable manifolds of the orbits on the attractor.

Thus, we get the first example of an autonomous distributed system with a uniformly hyperbolic attractor. As believed, it revives the old hope that such attractors may be relevant to some cases of complex dynamics of spatially extended systems (like the hydrodynamic turbulence). We can assume that the considered system may be implemented in electronics based on a kind of nonlinear transmission line. Attractiveness of systems with uniformly hyperbolic attractors in a frame of possible practical application of chaos is determined by their structural stability, or robustness: the generated chaos is insensitive to variations of parameters, imperfection of fabrication, technical fluctuations in the system, etc.

*This work was supported by RFBR grant No 11-02-91334 and DFG grant No PI 220/14-1. V.P.K. acknowledges support from DAAD in a frame of the program Forschungsstipendien für Doktoranden und Nachwuchswissenschaftler.*